\documentclass[conference]{IEEEtran}

\usepackage{cite}

\usepackage{microtype}

\usepackage{amsmath,amssymb}

\usepackage{graphicx}
\usepackage{capt-of}
\usepackage{float}

\usepackage{multirow}
\usepackage{makecell}
\usepackage{booktabs}
\usepackage{adjustbox}

\usepackage{xcolor}
\usepackage{pifont}
\usepackage{xspace}

\usepackage[colorlinks=true,
            citecolor=blue,
            linkcolor=blue]{hyperref}

\definecolor{darkgreen}{rgb}{0.0,0.5,0.0}
\definecolor{normalpurple}{rgb}{0.5,0.0,0.5}

\AtBeginDocument{
  \setlength{\abovedisplayskip}{3.5pt plus 1pt minus 1pt}
  \setlength{\belowdisplayskip}{3.5pt plus 1pt minus 1pt}
  \setlength{\abovedisplayshortskip}{2pt plus 1pt minus 1pt}
  \setlength{\belowdisplayshortskip}{2.5pt plus 1pt minus 1pt}
}

\hypersetup{pdftitle={ABLE: Anchor-Based Linear Equivalence for Reconstructing DRAM Address Mappings}}

\begin{document}

\title{ABLE: Anchor-Based Linear Equivalence for\\ Reconstructing DRAM Address Mappings}

\author{\IEEEauthorblockN{Jiajie Wen\IEEEauthorrefmark{1},
Xin Xin\IEEEauthorrefmark{2},
Hengyue Kang\IEEEauthorrefmark{1},
Jian Chen\IEEEauthorrefmark{3},\\
Mingyang Liu\IEEEauthorrefmark{1},
Congming Gao\IEEEauthorrefmark{1}, and
Jiwu Shu\IEEEauthorrefmark{1}}
\IEEEauthorblockA{\IEEEauthorrefmark{1}Xiamen University, Xiamen, China \quad
\IEEEauthorrefmark{2}University of Central Florida, Orlando, FL, USA\\
\IEEEauthorrefmark{3}Tsinghua University, Beijing, China\\
\{wenjiajie, mingyang\}@stu.xmu.edu.cn, xin.xin@ucf.edu, \{kanghengyue, gaocm92\}@gmail.com,\\
jian-che21@mails.tsinghua.edu.cn, jwshu@xmu.edu.cn\vspace{-12pt}}
}

\maketitle

\begin{abstract}
Understanding how physical addresses map to DRAM bank coordinates underpins Rowhammer exploitation and memory-security analysis, yet modern CPUs leave their XOR bank functions undocumented. Existing reverse-engineering tools struggle along three axes: fixed pairwise bit-flip probes miss single-bit functions and cross-function bit sharing, global linear-system solvers break under residual timing false positives, and brute-force mask search incurs exponential overhead with large memory footprints. We present ABLE (\emph{Anchor-Based Linear Equivalence}), a framework that reframes bank-mapping reconstruction via $\mathbb{F}_2$ linear equivalence. Observing that any partition-preserving basis is functionally identical to hardware bank indices, ABLE targets a reduced basis where each independent function holds a unique non-shared anchor bit. After filtering non-bank bits and grouping symmetric bits, ABLE selects an anchor basis and assigns remaining bits via controlled joint bit flipping stabilized by majority voting over independent conflict pairs. Across 12 Intel and AMD DDR4/DDR5 setups, ABLE reaches a 99\% mean success rate (94--100\%) in 2--15\,s without huge pages, running 12--27$\times$ faster than DARE.
\end{abstract}

\section{Introduction}
DRAM acts as the primary main memory in contemporary computer systems, making its security a critical concern~\cite{yauglikcci2021blockhammer}. Recent work has demonstrated increasingly sophisticated Rowhammer attacks, such as Blacksmith~\cite{Jattke2022BLACKSMITH}, ZenHammer~\cite{jattke2024zenhammer}, and $\rho$Hammer~\cite{chen2025rhohammer}.
Because attackers cannot address DRAM cells directly, they first need to determine how physical addresses map to DRAM locations. Accordingly, the reliability and precision of such attacks depend heavily on pinpointing target DRAM coordinates. Yet mainstream CPU vendors, notably Intel and AMD, no longer publicly document their bank address functions. Accurately and rapidly reconstructing DRAM address mappings has therefore become an essential prerequisite for Rowhammer research.

Although substantial progress has been made in reverse-engineering DRAM address mappings, existing methods still face three key limitations.
1) Brute-force search methods~\cite{Pessl2016drama,Wang2020DRAMDig,jattke2024zenhammer} first group same-bank addresses and then test candidate XOR combinations of physical address bits. However, candidate masks grow exponentially with the number of bits, making enumeration expensive, while gathering sufficiently informative address groups demands a large memory footprint.
2) The linear-algebraic method~\cite{plin2025knock} samples row-conflict pairs and solves a system of linear equations over $\mathbb{F}_2$. Yet false positives in conflict detection introduce invalid constraints that undermine algebraic recovery.
3) The structured pairwise testing method~\cite{chen2025rhohammer} probes relationships among address bits via controlled pairwise bit flips. Nevertheless, its predefined tests cannot resolve all structural relationships among address bits.

To overcome these limitations, we present ABLE (Anchor-Based Linear Equivalence), an anchor-guided method for efficient and reliable reconstruction of equivalent DRAM address mappings.
Our key observation is that recovering the exact hardware bank address functions is unnecessary, because any equivalent mapping that preserves the original bank partition suffices.
By exploiting the equivalence of linear bank mappings, ABLE constructs a simpler representation in which every independent bank function possesses a dedicated non-shared bit, called an anchor. This formulation turns mapping reconstruction from exhaustive XOR enumeration or solving noisy global constraints into a structured, anchor-guided procedure.
Concretely, ABLE selects independent anchor bits and systematically determines how the remaining address bits participate via controlled bit flipping.
To improve robustness against measurement errors, each candidate bit combination is evaluated across multiple independently sampled row-conflict pairs and decided by majority voting. Across our evaluated configurations, ABLE attains a 99\% average success rate (94--100\% per configuration), finishes in 2--15\,s, and achieves a 12--27$\times$ speedup over DARE~\cite{jattke2024zenhammer}.

Overall, we make the following contributions:
\begin{itemize}\setlength{\itemsep}{0.5pt}\setlength{\topsep}{1.5pt}\setlength{\parsep}{0pt}
    \item We present ABLE, a linear-equivalence paradigm for reconstructing equivalent DRAM address mappings.
    \item We design an anchor-guided method that combines linear transformations, controlled bit flipping, and majority voting for fast and noise-resilient mapping reconstruction.
    \item We evaluate ABLE across 12 Intel and AMD configurations with DDR4 and DDR5 memory, achieving a 99\% average success rate, completing reconstruction in 2--15\,s, and running 12--27$\times$ faster than DARE without requiring huge pages.
\end{itemize}

\section{Background and Related Work}
\subsection{DRAM Organization}
\begin{figure}[t]
\centering
\includegraphics[width=0.98\columnwidth]{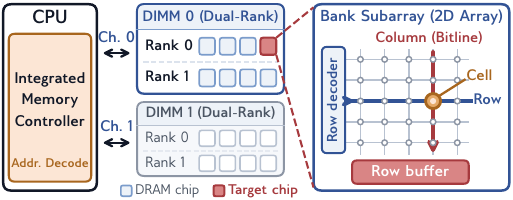}
\caption{Organization of a two-channel DRAM subsystem. The callout illustrates a 2D bank subarray inside one chip and the cell addressed by a row and column.}
\label{fig:DRAMOrg}
\end{figure}

Dynamic Random Access Memory (DRAM) provides main memory in modern computing platforms and is commonly packaged as Dual Inline Memory Modules (DIMMs). As illustrated in Fig.~\ref{fig:DRAMOrg}, the processor communicates with DIMMs over memory channels. Each DIMM comprises one or more ranks, and each rank is divided into multiple banks formed as two-dimensional cell arrays indexed by rows and columns. Upon a memory request, the integrated memory controller decodes the physical address to select the target channel, DIMM, rank, and bank, then activates the target row and accesses the requested column~\cite{jacob2008memory,deng2011memscale,kim2012case}.

\subsection{DRAM Address Mapping and Reverse Engineering}\label{sec:reverse_engineer}
\begin{figure}[t]
\centering
\includegraphics[width=0.98\columnwidth]{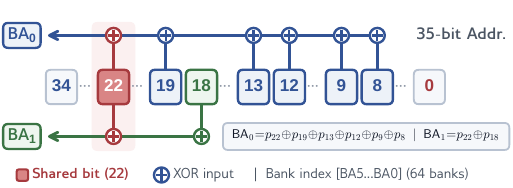}
\caption{Two XOR bank address functions on the 64-bank Comet Lake platform (35-bit physical address), where bit 22 is shared by $BA_0$ and $BA_1$.}
\label{fig:mappingExample}
\end{figure}

DRAM address mapping converts a physical address into bank, row, and column coordinates to locate a target memory cell. While row and column coordinates are determined directly by their corresponding physical address bits, the bank coordinate is computed by bank address functions (BFs), each typically implemented as XOR combinations of selected physical address bits, to spread contiguous physical addresses across banks and enable bank-level parallelism~\cite{zhang2000permutation,lin2001reducing,kaseridis2011minimalist}. Fig.~\ref{fig:mappingExample} shows two BFs on a Comet Lake (i7-10700) platform equipped with two DDR4 DIMMs. Here, a 35-bit physical address maps to a 6-bit bank coordinate $[BA_5, BA_4, BA_3, BA_2, BA_1, BA_0]$ indexing 64 banks; we refer to this complete bank coordinate as the \emph{bank index}. Each bit $BA_n$ is generated by a corresponding BF: for instance, $BA_0$ XORs physical address bits 8, 9, 12, 13, 19, and 22, whereas $BA_1$ XORs bits 18 and 22. In general, addressing $N$ banks requires at least $\log_2 N$ bank address bits, each governed by a corresponding BF.

On a given platform, the physical-to-DRAM mapping dictates which channel, DIMM, rank, bank, row, and column each physical address selects. Because vendors rarely document these mappings and vary them across platforms~\cite{yun2014palloc}, reconstructing them accurately is valuable for memory access optimization~\cite{Park2013regularities,Liu2014GoingVertical,Tang2016Improving,ghasempour2016dream}, write-latency reduction~\cite{Vittal2026BARD}, and hardware vulnerability studies such as Rowhammer~\cite{Kim2014disturbance,Xiao2016CrossVM,Cheng2021CATTmew,Kang2024SledgeHammer,deRidder2025Posthammer}. Consequently, numerous techniques have been proposed to reverse-engineer DRAM address mappings~\cite{Barenghi2018Softwareonly,Gerlach2024HashRecovery,Pessl2016drama,Wang2020DRAMDig,jattke2024zenhammer,chen2025rhohammer,plin2025knock,Jattke2025McSee}.

\noindent\textbf{Row-Buffer Timing Side Channel:}
Every DRAM bank includes a row buffer holding the currently open row. Servicing a memory request first \textit{activates} the target row into this buffer~\cite{seshadri2013rowclone,chang2014improving}. Subsequent accesses to the already open row yield a low-latency \textit{row-buffer hit}, whereas targeting another row in the same bank requires precharging the current row before activating the new one, causing a significantly higher-latency \textit{row-buffer conflict}~\cite{mutlu2008parallelism,rixner2000memory,kim2014bounding}. Prior work widely exploits this timing gap as a side channel to identify address pairs residing in different rows of the same bank~\cite{Pessl2016drama,Wang2020DRAMDig,jattke2024zenhammer,chen2025rhohammer,plin2025knock}. Concretely, we repeatedly alternate memory accesses between two randomly selected addresses: persistently high latency indicates that the two addresses map to different rows within the same bank, whereas low latency implies distinct banks because two randomly chosen addresses rarely fall in the same row.

\subsection{Related Work}\label{sec:related_work}
Brute-force search methods~\cite{Pessl2016drama,Wang2020DRAMDig,jattke2024zenhammer} group same-bank physical addresses via row-buffer conflict measurements and then enumerate candidate XOR combinations of physical address bits, retaining candidates whose outputs remain constant within each group while varying across groups. The linear-algebraic method~\cite{plin2025knock} formulates XOR bank-function recovery over $\mathbb{F}_2$, builds a pairwise XOR-difference matrix from sampled row-conflict pairs, and derives bank functions from its null space. The structured pairwise testing method~\cite{chen2025rhohammer} combines controlled pairwise bit flipping with row-conflict measurements to uncover relationships among physical address bits: it first uses pairwise tests to find bank functions containing row bits and then extends testing to larger bit combinations to recover functions without row bits.

\section{Motivation}\label{sec:motiva}
\subsection{Limitations of Prior Methods}
\begin{figure}[t]
\centering
\begin{minipage}[t]{0.530\columnwidth}
\centering
\vspace{0pt}
\includegraphics[width=\linewidth]{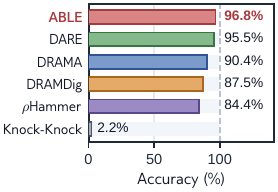}
\captionof{figure}{Proportion of selected high-latency address pairs that map to the same bank under the reference mapping, by method.}
\label{fig:compare_1}
\end{minipage}\hfill
\begin{minipage}[t]{0.445\columnwidth}
\centering
\vspace{0pt}
\includegraphics[width=\linewidth]{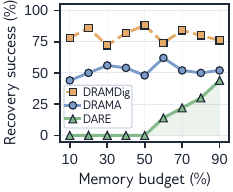}
\captionof{figure}{Mapping reconstruction success rate across memory budgets for DRAMA and DRAMDig on Comet Lake (16\,GiB DDR4) and DARE on Rocket Lake (32\,GiB DDR4).}
\label{fig:compare_2}
\end{minipage}
\end{figure}

\noindent\textbf{False Positives in Sampled Constraints:}
Elevated access latency does not always stem from a genuine row-buffer conflict. Fig.~\ref{fig:compare_1} evaluates the accuracy of row-conflict pair selection across multiple Intel and AMD desktop platforms against previously validated reference bank address functions used to determine whether two addresses belong to the same bank. Despite repeated timing trials, at least 2\% of selected high-latency pairs mapped to different banks under this reference. Moreover, these pairs continued to exhibit high latency across different system states, possibly due to internal DRAM remapping mechanisms. Although Knock-Knock employs subsampling and majority voting to mitigate measurement noise, residual false positives still inject erroneous constraints, causing its algebraic recovery to fail. Because a genuine row-conflict pair must produce identical outputs across all bank address functions, pairs yielding different outputs cannot represent genuine row conflicts. When we filter out these invalid pairs using the reference mapping, Knock-Knock successfully recovers the mapping, confirming that residual false positives undermine its algebraic inference; however, such filtering is impractical because the reference mapping is unavailable beforehand.

\noindent\textbf{Enumeration Cost and Sample Coverage:}
Brute-force methods, including DRAMA~\cite{Pessl2016drama}, DRAMDig~\cite{Wang2020DRAMDig}, and DARE~\cite{jattke2024zenhammer}, evaluate candidate functions over sampled address groups. Without structural restrictions, brute-force search must enumerate $2^k-1$ nonzero XOR masks over $k$ candidate physical address bits, incurring an exponentially growing search space. In addition, the sampled addresses must exhibit sufficient variation across the relevant physical address bits. Fig.~\ref{fig:compare_2} reports reconstruction success rates across memory budgets for DRAMA and DRAMDig on a Comet Lake platform with 16\,GiB DDR4, and for DARE on a Rocket Lake platform with 32\,GiB DDR4. On Comet Lake, DRAMA and DRAMDig show little change in success rate as the memory budget grows, whereas DARE's success rate on Rocket Lake improves with a larger memory budget. Brute-force search methods therefore face two key bottlenecks: exponential candidate enumeration and insufficient coverage of informative physical address samples.

\noindent\textbf{Incomplete Bit-Dependency Inference:}
Structured bit-flip testing avoids exhaustive enumeration of candidate functions, yet captures only the structural relationships covered by its predefined bit-flip tests. Whenever jointly flipping two physical address bits preserves the bank index, those two bits participate in the exact same bank address functions. Pairwise measurements thus identify bits sharing identical participation patterns. However, a matched pair may jointly participate in multiple bank functions and therefore does not necessarily isolate an individual function. Moreover, single-bit functions cannot be identified through pairwise tests alone, because their sole participating bit has no matching partner. Furthermore, shared bits introduce cross-function dependencies that pairwise measurements~\cite{chen2025rhohammer} cannot fully characterize, while exhaustively exploring higher-order bit combinations to resolve these dependencies is impractical.

\subsection{Reconsidering the Reconstruction Objective}
These limitations highlight three core challenges in reconstructing bank address functions: residual false positives corrupt algebraic constraints, brute-force search incurs exponential enumeration costs and requires sufficient physical-address coverage, and predefined pairwise measurements cannot fully resolve shared-bit dependencies or detect single-bit functions. Together, these observations indicate that reconstruction difficulty depends not only on measurement quality and search cost, but also on the structure of the target representation. For bank-mapping reconstruction, the essential information is whether two physical addresses belong to the same bank rather than their numerical bank indices, motivating us to reconsider the reconstruction target itself.

Because bank address functions are linear over $\mathbb{F}_2$~\cite{vandierendonck2005xor}, we can linearly transform their representations while preserving the original bank partition~\cite{plin2025knock}. Shared bits participate in multiple functions, making their effects on individual functions hard to distinguish through bit flipping, whereas single-bit functions have no matching partners for pairwise measurements. Choosing a suitable representation is therefore crucial: simpler bit-sharing relationships reduce the patterns that controlled bit flipping needs to distinguish. Can we exploit these linear transformations to find an equivalent representation that enables systematic reconstruction via controlled bit flipping?

\section{ABLE Design}
\subsection{Equivalent Linear Representation}
Consistent with prior work, we assume that bank address functions can be modeled as linear functions over $\mathbb{F}_2$~\cite{rau1991pseudo,Pessl2016drama,plin2025knock}, possibly after accounting for platform-specific address remapping.

\noindent\textbf{Linear Model of Bank Mapping:}
Let $n$ denote the number of physical address bits, and let $f_1,\ldots,f_m$ represent the original bank address functions. Writing a physical address as a binary column vector $\mathbf{p}= (p_1,\ldots,p_n)^{\mathsf T} \in\mathbb{F}_2^n$, we encode these functions as a binary coefficient matrix $\mathbf{M}\in\mathbb{F}_2^{m\times n}$ over the binary field $\mathbb{F}_2=\{0,1\}$, where row $i$ represents function $f_i$ and column $j$ corresponds to physical address bit $p_j$:
\[
    M_{ij}=
    \begin{cases}
        1, & \text{if } f_i \text{ includes } p_j,\\
        0, & \text{otherwise}.
    \end{cases}
\]
The output $b_i$ of each function $f_i$ is computed as
\[
    b_i=f_i(\mathbf{p})=
    \bigoplus_{j=1}^{n}\left(M_{ij}\land p_j\right),
    \qquad 1\le i\le m.
\]
Here, XOR and AND correspond to addition and multiplication over $\mathbb{F}_2$, respectively. Stacking the outputs of all $m$ functions yields the original bank mapping:
\begin{equation}
    F(\mathbf{p})=\mathbf{M}\mathbf{p}.
    \label{eq:linear_bank_mapping}
\end{equation}

\noindent\textbf{Equivalent Bank Mappings:}
Consider a candidate linear mapping $G(\mathbf{p})=\mathbf{B}\mathbf{p}$, where $\mathbf{B}\in\mathbb{F}_2^{k\times n}$. We define $F$ and $G$ to be equivalent, denoted by $F\sim G$, if for all $\mathbf{p},\mathbf{q}\in\mathbb{F}_2^n$,
\begin{equation}
    F(\mathbf{p}) = F(\mathbf{q})
    \quad\Longleftrightarrow\quad
    G(\mathbf{p}) = G(\mathbf{q}).
    \label{eq:bank_mapping_equivalence}
\end{equation}
The forward implication ensures that addresses assigned to the same bank by $F$ produce identical outputs under $G$, while the reverse implication ensures that addresses assigned to different banks by $F$ also yield distinct outputs under $G$. Together, these two conditions guarantee that $F$ and $G$ induce the exact same partition of the physical address space, even though their output labels (i.e., bank identifiers) may differ.

\noindent\textbf{Anchor-Bit Representation:}
We aim to construct an equivalent representation in which each resulting bank address function contains at least one non-shared bit. This representation serves two goals:
\ding{182} removing redundant functions and \ding{183} ensuring that every function contains at least one non-shared bit.
Concretely, we first discard linearly dependent functions and then reduce the remaining function matrix to reduced row echelon form.

Let $r=\operatorname{rank}(\mathbf{M})$ denote the number of linearly independent functions. Reordering the rows of $\mathbf{M}$ yields $\mathbf{M}=\begin{pmatrix}\mathbf{B}\\\mathbf{C}\end{pmatrix}$, where $\mathbf{B}$ contains $r$ linearly independent rows and $\mathbf{C}$ holds the remaining dependent rows. Because the rows of $\mathbf{C}$ are linear combinations of the rows of $\mathbf{B}$, there exists a matrix $\mathbf{D}$ such that $\mathbf{C}=\mathbf{D}\mathbf{B}$, allowing the original mapping to be written as
\[
\begin{aligned}
    F(\mathbf{p})
    &=
    \begin{pmatrix}
        \mathbf{B}\\
        \mathbf{C}
    \end{pmatrix}\mathbf{p}
    &=
    \begin{pmatrix}
        \mathbf{B}\mathbf{p}\\
        \mathbf{D}\mathbf{B}\mathbf{p}
    \end{pmatrix}.
\end{aligned}
\]
Setting $G(\mathbf{p})=\mathbf{B}\mathbf{p}$, we can express $F$ as
\[
    F(\mathbf{p})=
    \begin{pmatrix}
        G(\mathbf{p})\\
        \mathbf{D}G(\mathbf{p})
    \end{pmatrix}.
\]
Because $F(\mathbf{p})$ is completely determined by $G(\mathbf{p})$, any two physical addresses $\mathbf{p}$ and $\mathbf{q}$ satisfy
\[
    F(\mathbf{p})=F(\mathbf{q})
    \quad\Longleftrightarrow\quad
    G(\mathbf{p})=G(\mathbf{q}).
\]

All linearly dependent functions can be expressed as XOR combinations of the linearly independent ones, so their outputs are fixed once the independent function outputs are fixed. Equality of the complete function outputs is therefore governed solely by the $r$ independent functions, making them sufficient to preserve the original bank partition. Consequently, for a DRAM subsystem with $N$ physical banks, we have $2^r=N$, because each of the $r$ independent bank address functions produces one independent Boolean output bit.

\begin{figure}[t]
\centering
\includegraphics[width=0.98\columnwidth]{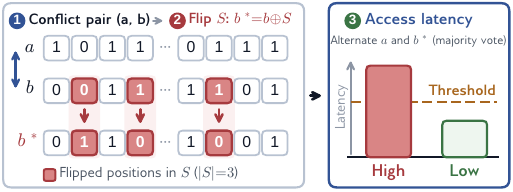}
\caption{Controlled bit-flipping primitive ($|S|=3$): (1) select a conflict pair $(a,b)$, (2) flip the shaded positions in $S$ to form $b^*=b\oplus S$, and (3) alternate accesses between $a$ and $b^*$ to test latency against a threshold via majority voting.}
\label{fig:flipping}
\end{figure}

Starting from the full-row-rank matrix $\mathbf{B}$, we apply Gauss--Jordan elimination over $\mathbb{F}_2$ to obtain its reduced row echelon form $\mathbf{R}$~\cite{jeannerod2013rank}.
Because $\mathbf{B}$ has full row rank, there exists an invertible matrix $\mathbf{U}\in\mathbb{F}_2^{r\times r}$ such that $\mathbf{R}=\mathbf{U}\mathbf{B}$. Letting $H(\mathbf{p})=\mathbf{R}\mathbf{p}$ denote the row-reduced mapping, the invertibility of $\mathbf{U}$ ensures that for any two physical addresses $\mathbf{p}$ and $\mathbf{q}$,
\[
\begin{aligned}
    H(\mathbf{p})=H(\mathbf{q})
    &\iff \mathbf{U}\mathbf{B}\mathbf{p}
          =\mathbf{U}\mathbf{B}\mathbf{q}\\
    &\iff \mathbf{B}\mathbf{p}
          =\mathbf{B}\mathbf{q}\\
    &\iff G(\mathbf{p})=G(\mathbf{q}).
\end{aligned}
\]

Because $\mathbf{R}$ is in reduced row echelon form, every pivot column contains a single $1$ and zeros in all other rows. Each row of $\mathbf{R}$ therefore possesses at least one non-shared bit, which serves as an anchor bit for the corresponding function. Over $\mathbb{F}_2$, Gauss--Jordan elimination consists of elementary row operations, including swapping rows and XORing one row with another. These operations correspond to reordering functions and replacing a function with the XOR of itself and other functions, thereby preserving equality relationships among mapping outputs even though numerical output labels may change.

\begin{figure}[t]
\centering
\includegraphics[width=0.98\columnwidth]{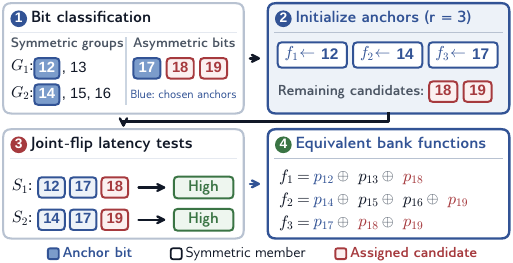}
\caption{Selecting anchors and assigning bits for three bank functions ($r=3$), where $p_j$ denotes physical address bit $j$.}
\label{fig:anchor}
\end{figure}

\subsection{Controlled Bit-Flipping Primitive}
ABLE builds its probing primitive by combining controlled bit flipping with row-buffer conflict measurements. As depicted in Fig.~\ref{fig:flipping}, we start from a row-conflict pair $(a,b)$ within the same bank. Given a target bit set $S$, we flip the corresponding bits in $b$ to obtain $b^*=b\oplus S$, then alternate accesses between $a$ and $b^*$ and measure their access latency. If high access latency persists, flipping $S$ preserves the bank index; otherwise, the bank index may have changed. By varying $S$, ABLE probes how individual physical address bits and their combinations affect bank indexing, forming the basis for subsequent bit detection. To mitigate false positives and latency measurement noise, each candidate bit set $S$ is evaluated across multiple independently sampled conflict address pairs and decided via majority voting.

\begin{table*}[t]
\caption{Reconstructed DRAM Address Mappings Across 12 Intel and AMD Desktop Configurations.}
\label{tab:dram_address_mappings}
\centering
\footnotesize
\setlength{\tabcolsep}{12pt}
\renewcommand{\arraystretch}{0.8}
\setlength{\aboverulesep}{0.5pt}
\setlength{\belowrulesep}{0.5pt}

\begin{adjustbox}{max width=\textwidth}
\begin{tabular}{@{}ccc@{\hspace{6pt}}c@{\hspace{6pt}}ccc@{}}
\toprule[1.0pt]

\multirow{2}{*}{\textbf{Config. ID}}
& \multirow{2}{*}{\textbf{Microarch.}}
& \multicolumn{3}{c}{\textbf{DRAM}}
& \multirow{2}{*}{\textbf{Bank Address Functions}}
& \multirow{2}{*}{\textbf{Pure Column Bits}} \\
\cmidrule(lr){3-5}
& & \textbf{Type} & \textbf{Size} & \textbf{Banks} & & \\

\midrule[0.6pt]

\multirow{2}{*}{1} & Comet Lake
& \multirow{2}{*}{DDR4}
& \multirow{2}{*}{16\,GiB}
& \multirow{2}{*}{32}
& (6, 13), (14, 18),
& \multirow{2}{*}{0$\sim$5, 7$\sim$12} \\
& i7-10700 & & &
& (15, 19), (16, 20), (17, 21)
& \\

\midrule[0.3pt]

\multirow{2}{*}{2} & Comet Lake
& \multirow{2}{*}{DDR4}
& \multirow{2}{*}{32\,GiB}
& \multirow{2}{*}{64}
& (7, 14), (15, 19), (16, 20),
& \multirow{2}{*}{0$\sim$6, 10, 11} \\
& i7-10700 & & &
& (17, 21), (18, 22), (8, 9, 12, 13, 19, 22)
& \\

\midrule[0.6pt]

\multirow{2}{*}{3} & Rocket Lake
& \multirow{2}{*}{DDR4}
& \multirow{2}{*}{16\,GiB}
& \multirow{2}{*}{32}
& (6, 13), (14, 18),
& \multirow{2}{*}{0$\sim$5, 7$\sim$12} \\
& i7-11700 & & &
& (15, 19), (16, 20), (17, 21)
& \\

\midrule[0.3pt]

\multirow{2}{*}{4} & Rocket Lake
& \multirow{2}{*}{DDR4}
& \multirow{2}{*}{32\,GiB}
& \multirow{2}{*}{64}
& (6, 14), (16, 20), (15, 19, 27, 30, 33), (8, 9, 12, 13, 18, 19),
& \multirow{2}{*}{0$\sim$5, 7, 10, 11} \\
& i7-11700 & & &
& (18, 22, 23, 26, 29, 32), (17, 21, 24, 25, 28, 31, 34)
& \\

\midrule[0.6pt]

\multirow{2}{*}{5} & Alder Lake
& \multirow{2}{*}{DDR4}
& \multirow{2}{*}{32\,GiB}
& \multirow{2}{*}{64}
& (16, 20), (10, 12, 14), (9, 10, 13, 19), (15, 19, 27, 30, 33),
& \multirow{2}{*}{0$\sim$8, 11} \\
& i7-12700 & & &
& (17, 21, 24, 25, 28, 31, 34), (18, 22, 23, 26, 29, 32)
& \\

\midrule[0.3pt]

\multirow{2}{*}{6} & Alder Lake
& \multirow{2}{*}{DDR5}
& \multirow{2}{*}{32\,GiB}
& \multirow{2}{*}{128}
& (8, 13, 20), (9, 14, 20), (9, 10, 13, 19), (18, 23, 27, 31),
& \multirow{2}{*}{0$\sim$7, 11, 12} \\
& i7-12700 & & &
& (15, 19, 24, 28, 32), (16, 21, 25, 29, 32), (17, 22, 26, 30, 34)
& \\

\midrule[0.6pt]

\multirow{2}{*}{7} & Raptor Lake
& \multirow{2}{*}{DDR4}
& \multirow{2}{*}{32\,GiB}
& \multirow{2}{*}{64}
& (16, 20), (10, 12, 14), (9, 10, 13, 19), (15, 19, 27, 30, 33),
& \multirow{2}{*}{0$\sim$8, 11} \\
& i7-14700 & & &
& (17, 21, 24, 25, 28, 31, 34), (18, 22, 23, 26, 29, 32)
& \\

\midrule[0.3pt]

\multirow{2}{*}{8} & Raptor Lake
& \multirow{2}{*}{DDR5}
& \multirow{2}{*}{32\,GiB}
& \multirow{2}{*}{128}
& (8, 13, 20), (9, 14, 20), (9, 10, 13, 19), (18, 23, 27, 31),
& \multirow{2}{*}{0$\sim$7, 11, 12} \\
& i7-14700 & & &
& (15, 19, 24, 28, 32), (16, 21, 25, 29, 32), (17, 22, 26, 30, 34)
& \\

\midrule[1.0pt]

\multirow{2}{*}{9} & Zen 4
& \multirow{2}{*}{DDR5}
& \multirow{2}{*}{16\,GiB}
& \multirow{2}{*}{64}
& (8, 20, 25, 30), (9, 21, 26, 31), (11, 19, 24, 29),
& \multirow{2}{*}{0$\sim$5, 7, 13$\sim$17} \\
& R5-7600X & & &
& (12, 22, 27, 32), (10, 18, 23, 28, 33), (6, 8, 9, 10, 11, 12)
& \\

\midrule[0.3pt]

\multirow{2}{*}{10} & Zen 4
& \multirow{2}{*}{DDR5}
& \multirow{2}{*}{32\,GiB}
& \multirow{2}{*}{128}
& (8), (9, 21, 26, 31), (10, 22, 27, 32), (12, 20, 25, 30),
& \multirow{2}{*}{0$\sim$5, 7, 14$\sim$18} \\
& R5-7600X & & &
& (13, 23, 28, 33), (11, 19, 24, 29, 34), (6, 9, 10, 11, 12, 13)
& \\

\midrule[0.6pt]

\multirow{2}{*}{11} & Zen 5
& \multirow{2}{*}{DDR5}
& \multirow{2}{*}{16\,GiB}
& \multirow{2}{*}{64}
& (8, 20, 25, 30), (9, 21, 26, 31), (11, 19, 24, 29),
& \multirow{2}{*}{0$\sim$5, 7, 13$\sim$17} \\
& R5-9600X & & &
& (12, 22, 27, 32), (10, 18, 23, 28, 33), (6, 8, 9, 10, 11, 12)
& \\

\midrule[0.3pt]

\multirow{2}{*}{12} & Zen 5
& \multirow{2}{*}{DDR5}
& \multirow{2}{*}{32\,GiB}
& \multirow{2}{*}{128}
& (8), (9, 21, 26, 31), (10, 22, 27, 32), (12, 20, 25, 30),
& \multirow{2}{*}{0$\sim$5, 7, 14$\sim$18} \\
& R5-9600X & & &
& (13, 23, 28, 33), (11, 19, 24, 29, 34), (6, 9, 10, 11, 12, 13)
& \\

\bottomrule[1.0pt]
\end{tabular}
\end{adjustbox}
\end{table*}

\subsection{Anchor-Guided Function Reconstruction}
\noindent\textbf{Non-Bank Bit Filtering:}
\ding{182}
Following prior work~\cite{Xiao2016CrossVM}, we first detect pure row bits by randomly picking a physical address and flipping each bit one at a time. If alternating accesses between the original and flipped addresses incurs high latency, the two addresses reside in distinct rows of the same bank, identifying the flipped bit as a pure row bit.
\ding{183}
Next, we set $S$ to one physical address bit at a time and evaluate each bit using our controlled bit-flipping primitive. If high access latency persists, flipping that bit does not alter the bank index, so it is classified as a non-bank bit; otherwise, it is classified as a bank bit. Excluding the identified pure row bits from the non-bank bits yields the pure column bits.

\noindent\textbf{Symmetric Bit Grouping:}
Physical address bits that participate in the exact same set of bank functions are termed symmetric bits, whereas a bank bit without a symmetric partner is termed asymmetric. When two-bit sets $S$ are tested one pair at a time via our controlled bit-flipping primitive, persistent high latency shows that jointly flipping the pair preserves the bank index and hence that both bits share identical participation patterns across all bank functions. Merging overlapping matched pairs forms symmetric bit groups.

\noindent\textbf{Selecting Anchors and Assigning Bits:}
Any set of $r$ linearly independent bank-bit columns can serve as an anchor basis, allowing every remaining physical address bit to be expressed uniquely as an XOR combination of the anchors and recovered via controlled bit flipping.
We build a candidate set by choosing one representative from each symmetric bit group together with all asymmetric bits, and select $r$ anchors from this set to initialize $r$ bank address functions. We prioritize group representatives and use asymmetric bits to fill any remaining slots. Fig.~\ref{fig:anchor} illustrates this procedure for $r=3$.

For every remaining candidate bit $p_j$, we apply the controlled bit-flipping primitive to test flipping $p_j$ jointly with subsets of the selected anchors. If high access latency persists for an anchor subset, the joint flip preserves the bank index; we then add $p_j$ to each bank function initialized by an anchor in that subset and proceed to the next candidate bit. If no anchor subset passes the latency test, $p_j$ cannot be incorporated, and we reject the current anchor set.

Once all candidate bits are assigned, each non-representative bit is added to the same bank functions as its group representative, yielding an equivalent representation of the original mapping. Following a rejection, we first try alternative combinations of asymmetric bits and then alternative combinations of group representatives. If all combinations with the current counts fail, we decrease the number of group representatives and increase the number of asymmetric bits while keeping $r$ anchors in total, repeating the search until a valid anchor set is found.

\section{Evaluation}
We evaluate ABLE across 12 Intel and AMD desktop configurations spanning DDR4 and DDR5 memory, as summarized in Table~\ref{tab:dram_address_mappings}. All experiments run on Ubuntu 26.04 LTS under default BIOS settings using 16\,GiB (single-DIMM) and 32\,GiB (dual-DIMM) memory configurations.

\subsection{Settings}
\noindent\textbf{Physical Address Acquisition:}
In line with prior studies~\cite{Pessl2016drama,Wang2020DRAMDig,jattke2024zenhammer,chen2025rhohammer,plin2025knock}, reconstruction is conducted offline with root privileges to read physical frame numbers (PFNs) via \texttt{/proc/self/pagemap}, from which we build an inverse lookup table mapping physical page addresses to virtual page addresses. Whenever bit flipping produces a target physical address, we query this table for its virtual address; if the target physical page is absent from the table, we skip the candidate and try another address. ABLE thus searches for accessible addresses satisfying the required bit-flip conditions.

\noindent\textbf{Bank Count Estimation and Mapping Validation:}
To estimate the bank count $\hat{N}$, we randomly sample physical addresses and retain an address only when it exhibits no row-buffer conflict with any previously retained address, treating each retained address as a distinct bank representative. Under roughly uniform sampling, once more than half of the banks are covered, new draws hit already covered banks more often than uncovered ones~\cite{boneh1997coupon}; we therefore stop after several consecutive draws fail to add a new representative and round the retained count up to the nearest power of two.

Each reconstructed mapping is validated on a separate set of conflict address pairs not used during recovery. Defining consistency as the fraction of validation pairs whose two addresses yield identical outputs across all recovered bank functions, a mapping passes validation when its consistency exceeds 96\% and its coefficient matrix attains the expected rank $\hat{r}=\log_2\hat{N}$, where the 96\% threshold accommodates false positives in row-buffer conflict detection.

\begin{figure}[t]
\centering
\includegraphics[width=0.98\columnwidth]{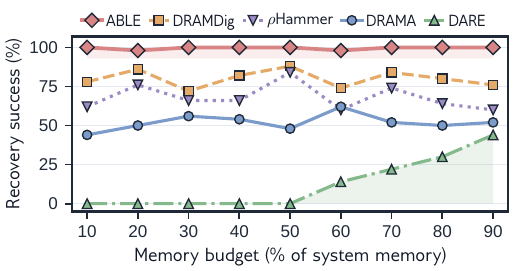}
\caption{Reconstruction success rate versus memory budget: ABLE and DARE on Rocket Lake (32\,GiB DDR4); DRAMA, DRAMDig, and $\rho$Hammer on Comet Lake (16\,GiB DDR4).}
\label{fig:memory_requirement}
\end{figure}

\subsection{Reconstructed DRAM Address Mappings}
As reported in Table~\ref{tab:dram_address_mappings}, ABLE reconstructs the complete bank address functions and identifies pure column bits across all 12 tested configurations. At a fixed bank count and DRAM configuration, newer Intel microarchitectures employ more complex bank functions involving more physical address bits: among the 64-bank, 32\,GiB DDR4 configurations, for instance, Rocket Lake, Alder Lake, and Raptor Lake span more physical address bits across their bank functions than Comet Lake. Shared bits appear in all evaluated 64- and 128-bank configurations but in none of the 32-bank configurations, while single-bit bank functions appear on the 128-bank Zen 4 and Zen 5 configurations.

\subsection{Performance Comparison}
We compare ABLE against existing tools along three dimensions: memory requirement, stability, and runtime efficiency. Each method is executed 50 times per configuration to measure its recovery success rate (the percentage of runs passing mapping validation) and mean runtime.

Knock-Knock is omitted from the performance plots because it did not recover valid bank functions on our tested Intel and AMD platforms. As shown in Section~\ref{sec:motiva}, Knock-Knock succeeded once sampled pairs were filtered with the reference mapping, which is unavailable during practical reverse engineering.

In the stability and efficiency evaluations, DARE is granted the maximum available huge pages, whereas ABLE allocates 15\% of physical memory via anonymous \texttt{mmap} without huge pages.

\noindent\textbf{Memory Requirement:}
Fig.~\ref{fig:memory_requirement} examines how memory budgets affect recovery on a Rocket Lake platform with 32\,GiB DDR4. ABLE maintains a near-100\% success rate across all tested budgets, showing negligible sensitivity to allocated memory size. By contrast, DARE depends heavily on the memory budget, failing completely below 60\% and reaching only 44\% success at 90\%. Because DRAMA and DRAMDig recover mappings only on 16\,GiB DDR4 Comet Lake and Rocket Lake configurations, we evaluate their budget sensitivity on Comet Lake (16\,GiB DDR4), where DRAMDig stays below 90\% and DRAMA remains near 50\%. On this same Comet Lake configuration, whose bank functions share no physical address bits, $\rho$Hammer attains 60--84\% success across the tested memory budgets.

\begin{figure}[t]
\centering
\includegraphics[width=0.98\columnwidth]{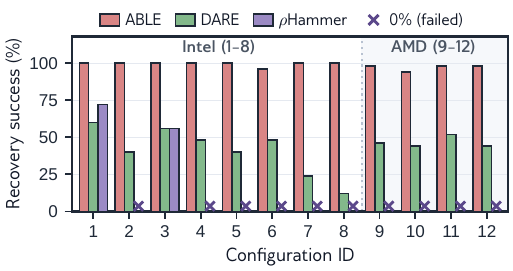}
\caption{Recovery success rate (percentage of 50 runs passing mapping validation) across the 12 Intel and AMD configurations in Table~\ref{tab:dram_address_mappings}. Crosses ($\times$) denote 0\% success.}
\label{fig:reverse_engineered_accuracy}
\end{figure}

\noindent\textbf{Stability:}
Fig.~\ref{fig:reverse_engineered_accuracy} compares recovery success rates across all 12 desktop configurations. ABLE is the sole evaluated method achieving 94--100\% success across every configuration. $\rho$Hammer succeeds only on configurations with neither shared bits nor single-bit functions, failing on the remaining 10 configurations (all four AMD configurations and six Intel configurations). DARE remains the strongest brute-force baseline, yet reaches only 12--60\% success across these platforms despite using the maximum available huge pages.

\begin{figure}[H]
\centering
\includegraphics[width=0.98\columnwidth]{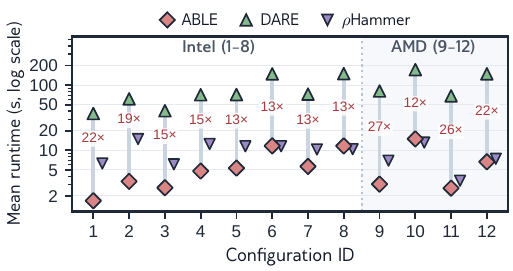}
\caption{Mean time to termination (log scale) over 50 runs per method across the 12 configurations in Table~\ref{tab:dram_address_mappings}, including runs that failed validation. Labels along each vertical stem indicate ABLE's speedup over DARE.}
\label{fig:reverse_engineered_cost}
\end{figure}

\noindent\textbf{Efficiency:}
Fig.~\ref{fig:reverse_engineered_cost} compares mean runtimes among ABLE, $\rho$Hammer, and DARE. ABLE reconstructs bank address functions within 2--15\,s across all 12 configurations. Although $\rho$Hammer terminates within 3--15\,s, it fails validation on 10 configurations. By contrast, DARE takes 37--173\,s, running 12--27$\times$ slower than ABLE.

Taken together, ABLE is the only evaluated method that simultaneously delivers a small memory footprint, consistently high success rates, and second-scale reconstruction times across all tested configurations.

\section{Conclusion}
Modern memory controllers employ intricate XOR bank functions with shared and single-bit terms that hinder existing reverse-engineering methods. ABLE resolves these barriers by targeting a partition-equivalent $\mathbb{F}_2$ basis whose non-shared anchor bits decouple reconstruction into majority-voted joint bit-flipping tests without exhaustive enumeration. Across 12 DDR4/DDR5 Intel and AMD setups, ABLE averages 99\% success within 2--15\,s without huge pages, outperforming DARE by 12--27$\times$.

\bibliographystyle{IEEEtran}
\bibliography{main}

\end{document}